\documentclass[
aps,
prl,
showpacs,
twocolumn
]{revtex4-1}
\usepackage{amsthm}
\theoremstyle{plain}
\newtheorem{theorem}{Theorem}%[section]
\newtheorem*{theorem*}{Theorem}

\newtheorem*{definition*}{Definition}

\newtheorem*{lemma*}{Lemma}

\usepackage{braket}
\usepackage{delarray}
\usepackage{here}

\usepackage{amsmath}
\usepackage{txfonts}

\usepackage{graphicx}
\usepackage{bm}
\usepackage{color}
\usepackage{ulem}

\usepackage{physics}

\usepackage{tikz}
\usepackage{lipsum}

\newcommand{\ba}{\begin{array}}
\newcommand{\ea}{\end{array}}
\newcommand{\bmat}{\left(\begin{array}}
\newcommand{\emat}{\end{array}\right)}
\newcommand{\no}{\nonumber}

\newcommand{\be}{\begin{eqnarray}}
\newcommand{\ee}{\end{eqnarray}}

\begin{document}
\title{Absence of Spin-Glass Order on Migdal--Kadanoff Hierarchical Lattices near Three Dimensions}
\author{Manaka Okuyama$^{1,2}$}
\author{Masayuki Ohzeki$^{1,3,4,5}$}
\affiliation{$^1$Graduate School of Information Sciences, Tohoku University, Sendai 980-8579, Japan}
\affiliation{$^2$School of Computing, Institute of Science Tokyo, Tokyo 152-8551, Japan}
\affiliation{$^3$Department of Physics, Institute of Science Tokyo, Tokyo 152-8551, Japan}
\affiliation{$^4$Research and Education Institute for Semiconductors and Informatics, Kumamoto University, Kumamoto 860-0862, Japan}
\affiliation{$^5$Sigma-i Co., Ltd., Tokyo 108-0075, Japan} %\\

\begin{abstract}

We derive a rigorous sufficient condition for the absence of spin-glass order in the Ising spin glass with symmetric binary couplings on Migdal--Kadanoff (MK) hierarchical lattices with even branching number. The key observation is that a single exact renormalization-group step creates zero effective bonds with positive probability, thereby reducing the problem to bond percolation on the corresponding hierarchical lattice. 
When the induced dilution exceeds the percolation threshold, both spin-glass order and stiffness are absent at all temperatures, including zero temperature. 
As a consequence, our criterion gives a rigorous proof that the Ising spin glass on the square lattice does not exhibit a spin-glass phase within the MK approximation. More unexpectedly, by choosing sufficiently large scale factors and branching numbers, we construct MK hierarchical lattices whose fractal dimensions are arbitrarily close to three from below but still exhibit no spin-glass order. 
This sharply contrasts with numerical estimates obtained for MK hierarchical lattices with relatively small scale factors and branching numbers, which placed the lower critical dimension near \(2.52\).

\end{abstract}
\date{\today}
\maketitle

%%%%%%%%%%%%%%%%%%%%%%%%%%%%%%%%%%%%%%%%%%%%%%%%%%%%%%%%%%%%%%%%%%%%%%%%%%%
%%%%%%%%%%%%%%%%%%%%%%%%%%%%%%%%%%%%%%%%%%%%%%%%%%%%%%%%%%%%%%%%%%%%%%%%%%%
%\section{Introduction}
\paragraph*{Introduction---}

While mean-field spin-glass models are now understood with remarkable mathematical precision~\cite{Parisi,Guerra,Talagrand}, finite-dimensional spin-glass models remain far less tractable~\cite{NS,NS2,NRS,Chatterjee,NRS2,NOO,NS3}.
Even basic questions, such as the existence or absence of a spin-glass phase in low dimensions, are not rigorously settled.
Numerical studies suggest the absence of a spin-glass phase at finite temperature in two dimensions and its presence in three dimensions~\cite{McMillan,BM2,McMillan2,BY,HY,HH,SC,KR,SP}.
The lower critical dimension is estimated to be \(d_{\rm lc}\simeq 2.5\)~\cite{MP,Boettcher,FPV}; however, no rigorous derivation of this picture is currently available.

Hierarchical lattices provide a useful setting in which aspects of finite-dimensional spin glasses can be studied using real-space renormalization-group (RG) methods.
In particular, Migdal--Kadanoff (MK) hierarchical lattices~\cite{Migdal,Kadanoff}, specified by a scale factor \(b\) and a branching number \(s\), have long provided a standard framework for applying such RG methods to finite-dimensional spin glasses~\cite{YS,MBK,SY,JCW,BM,CM,Gingras,ABD,CNC,ANC,AMMP,JK,DTB,DBMB,SM,NH,BKM,AB}.
Their fractal dimension, \(d=1+\log s/\log b\), can be varied by changing \(s\) and \(b\), while their recursive structure makes the RG transformation exact. 
This exact recursion enables numerical studies of very large systems.
Existing rigorous analyses, however, have largely been restricted to models with sufficiently large fractal dimensions~\cite{Gardner,CE,Kamphorst}, leaving the low-dimensional regime near the putative lower critical dimension poorly understood. 
Numerical studies of MK hierarchical lattices with relatively small \(s\) and \(b\) have estimated the lower critical dimension of the Ising spin glass to be \(d_{\rm lc}^{\rm MK}\simeq2.52\)~\cite{DTB}. 
By contrast, a scaling argument in the large-\(s\), large-\(b\) regime identifies \(d=3\) as marginal~\cite{Boettcher}, suggesting that the apparent lower-critical behavior may depend not only on the fractal dimension but also on the detailed hierarchical construction.

In this Letter, we introduce a simple and rigorous mechanism for excluding spin-glass order on MK hierarchical lattices with binary couplings.
Instead of iterating the full distributional RG equation, we perform a single exact RG step for each realization.
For a symmetric binary distribution and an even branching number \(s\), the renormalized interaction vanishes exactly with positive probability.
These zero effective bonds induce an exact bond dilution.
The problem of excluding spin-glass order can then be reduced to bond percolation on the corresponding hierarchical lattice.
This observation yields a rigorous sufficient condition for the absence of spin-glass order.
Whenever the induced dilution exceeds the hierarchical percolation threshold, the probability that the endpoints remain connected decays exponentially with their separation.
Consequently, the endpoint spin-glass correlation vanishes at any temperature, including zero temperature, in the thermodynamic limit.
Under the same condition, we also prove the absence of stiffness.

Several consequences follow immediately.
First, the criterion applies to the MK approximation of the Ising spin glass on the square lattice. 
Although the absence of a spin-glass phase in this approximation has long been expected~\cite{YS}, the present result provides, to the best of our knowledge, the first rigorous proof.
Second, and more unexpectedly, by taking both \(s\) and \(b\) sufficiently large, we construct MK hierarchical lattices that satisfy the no-order criterion even though their fractal dimensions can be made arbitrarily close to three from below.
This result contrasts sharply with the commonly quoted estimate \(d_{\rm lc}^{\rm MK}\simeq 2.52\)~\cite{DTB} and rigorously establishes the no-order side of the scaling prediction that \(d=3\) is marginal in the large-\(s\), large-\(b\) regime~\cite{Boettcher}.

We also show that the same dilution mechanism is not restricted to MK hierarchical lattices.
As an example, we apply it to the simplest self-dual hierarchical lattice, whose fractal dimension is \(d=\log 5/\log 2\simeq 2.32\)~\cite{GK,KG2,Nobre}.
Previous work~\cite{ON} based on graph duality, the replica method, and real-space RG arguments suggested the absence of a spin-glass phase at finite temperature on this lattice; here we rigorously establish the absence of both spin-glass order and stiffness at all temperatures.

\begin{figure}[!t]
\centering
\includegraphics[width=\columnwidth]{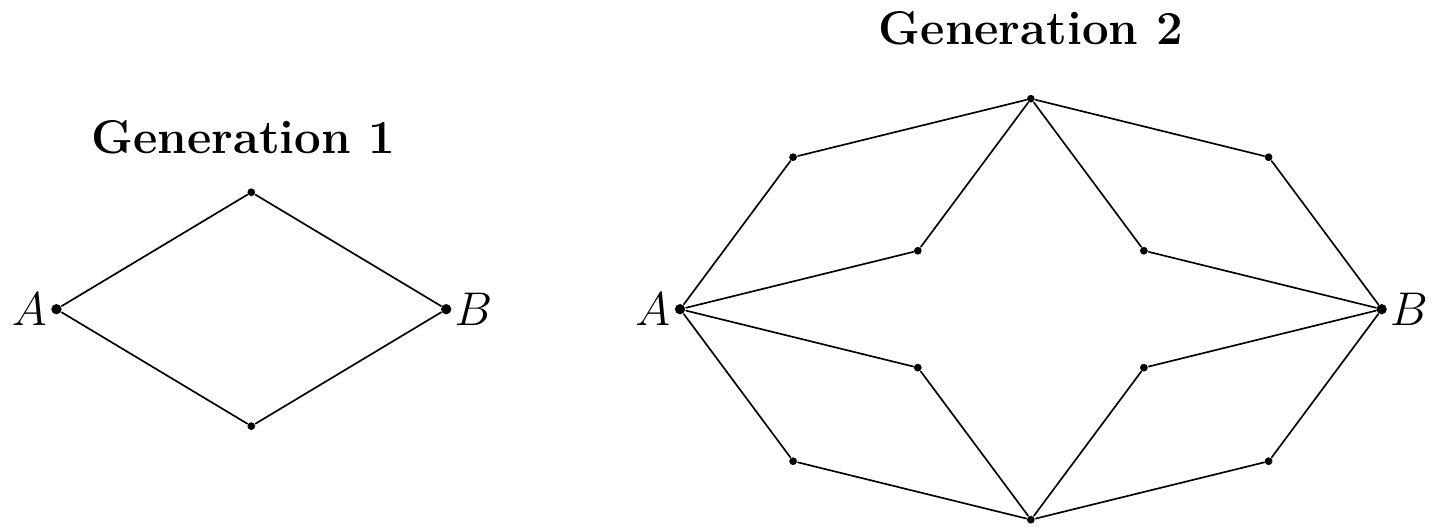}
\caption{First two generations of the MK hierarchical lattice for
\(s=b=2\). Each bond is replaced by two parallel chains of two bonds.}
\label{fig:mk}
\end{figure}

%%%%%%%%%%%%%%%%%%%%%%%%%%%%%%%%%%%%%%%%%%%%%%%%%%%%%%%%%%%%%%%
%\section{Absence of spin glass phase in Migdal--Kadanoff hierarchical lattices}
\paragraph*{Absence of spin-glass order on a class of Migdal--Kadanoff hierarchical lattices---}
Let \(b\ge 2\) and \(s\ge 1\) denote the scale factor and the branching number, respectively. 
The Migdal--Kadanoff (MK) hierarchical lattice is constructed recursively by replacing each edge with \(s\) parallel chains, each of length \(b\) (see Fig.~\ref{fig:mk} in the End Matter).
Generation \(0\) consists of a single edge, and the distance between the two endpoints \(A\) and \(B\) at generation \(n\) is $r_n=b^n$.
The fractal dimension of the lattice is $d=1+{\log s}/{\log b}$.
For a single generating unit, the Hamiltonian is
\begin{equation}
\mathcal H_{\rm unit}=-\sum_{\alpha=1}^{s}\sum_{k=0}^{b-1}J_{\alpha k}\,\sigma_{\alpha,k}\sigma_{\alpha,k+1},
\end{equation}
where each \(\sigma_{\alpha,k}=\pm1\) denotes an Ising spin, with \(\sigma_{\alpha,0}\equiv\sigma_L\) and \(\sigma_{\alpha,b}\equiv\sigma_R\).
The couplings are independent and identically distributed random variables drawn from the symmetric binary distribution, $P(J_{ij})
=
\delta(J_{ij}-1)/2
+
\delta(J_{ij}+1)/2$.
The Hamiltonian for the Ising spin-glass model on the MK hierarchical lattice at generation \(n\) is $H_n=-\sum_{(ij)\in E^{(n)}}J_{ij}\sigma_i\sigma_j$,
where \(E^{(n)}\) denotes the edge set at generation \(n\). We impose free boundary conditions at the two endpoints \(A\) and \(B\). The endpoint spin-glass correlation is defined by $\mathbb{E}\!\left[\langle \sigma_A\sigma_B\rangle_n^2\right]$,
where \(\langle\cdots\rangle_n\) denotes the thermal average with respect to \(H_n\), and \(\mathbb{E}[\cdots]\) denotes the disorder average. The absence of endpoint spin-glass order means
\be
\lim_{n\to\infty}
\mathbb{E}\!\left[\langle \sigma_A\sigma_B\rangle_n^2\right]  
=0 .\label{end-SG}
\ee
Note that if the endpoint spin-glass correlation vanishes, then the square of the conventional spin-glass order parameter,
\be
\frac{1}{|V^{(n)}|^2}
\sum_{i,j\in V^{(n)}}
\mathbb{E}\left[
\langle \sigma_i\sigma_j\rangle_n^2
\right],
\ee
also vanishes. Here, \(V^{(n)}\) denotes the vertex set at generation \(n\).
Therefore, the vanishing of the endpoint spin-glass correlation in Eq.~\eqref{end-SG} also rules out a nonzero conventional spin-glass order parameter.

Because of the recursive structure of the MK hierarchical lattice~\cite{KG}, summing exactly over the internal spins of a generating unit produces a single effective interaction \(J_{LR}\) between the two endpoint spins. 
At finite temperature, the exact RG equation~\cite{JK} is
\be
\beta J_{LR}&=&\sum_{\alpha=1}^{s} \tanh^{-1}\left( \prod_{k=0}^{b-1} \tanh(\beta J_{\alpha k})\right), \label{interaction-update}
\ee
where \(\beta\) denotes the inverse temperature.
Equivalently, the probability distribution of the effective coupling is updated according to the same equation, with equality understood in distribution.

In contrast to the usual approach, which repeatedly iterates this distributional RG equation, we use only a single exact RG step for each realization.
For the symmetric binary distribution, each branch contribution in Eq.~\eqref{interaction-update} has the same magnitude and an independent random sign. Hence, if \(s\) is even, the effective interaction \(J_{LR}\) vanishes exactly when the numbers of positive and negative branch contributions are equal. The probability of this event is
\be
p_{s,b}
=
\binom{s}{s/2}\frac{1}{2^s}. \label{psb}
\ee
For the symmetric binary distribution, this probability is independent of the scale factor \(b\). This independence is special to the symmetric case; as discussed below, \(p_{s,b}\) generally depends on \(b\) for asymmetric binary distributions.

After applying this single RG step to every elementary unit of the MK hierarchical lattice at generation \(n\), we obtain an MK hierarchical lattice at generation \(n-1\) in which each effective bond is absent with probability \(p_{s,b}\).
We now discard the values of the nonzero effective interactions and retain only their connectivity. 
If the endpoints \(A\) and \(B\) are disconnected after the zero effective bonds are removed, then the Hamiltonian decomposes into independent components containing \(A\) and \(B\), respectively.
Since there is no external field and the endpoint boundary conditions are free, this implies
\be
\langle \sigma_A\sigma_B\rangle_n=0 .
\ee
Therefore, the spin-glass correlation is bounded above by the probability that \(A\) and \(B\) remain connected through nonzero effective bonds.
This reduces the problem to bond percolation on the MK hierarchical lattice. 
For a single MK generating unit, the probability that no path connects its two endpoints is $[1 - (1 - p_{s,b})^b]^s$.
Thus, if the initial vacancy probability \(p_{s,b}\) satisfies
\be
p_{s,b}
<
\left[1-(1-p_{s,b})^b\right]^s , \label{sufficient-condition}
\ee
then the vacancy probability increases under the hierarchical percolation recursion. 
Equivalently, the endpoint connectivity probability decreases under iteration.
A detailed proof, including the fixed-point analysis of the percolation recursion and the derivation of the exponential bound, is given in the End Matter.
This yields the following sufficient condition for the absence of spin-glass order.

%%%%%%%%%%%%%%%%%%%%%%%%%%%%%%%%%%%%%%%%%%%%%%%%%%%%%%%%%%%%%%%
\begin{theorem}
Let \(s\) be even. Suppose that \(p_{s,b}\) satisfies condition~\eqref{sufficient-condition}. Then there exist constants \(C>0\) and \(c>0\), independent of \(n\), such that $\mathbb{E}\!\left[
\langle \sigma_A\sigma_B\rangle_n^2
\right]
\le
C \exp(-c r_n)$.
In particular,
\be
\lim_{n\to\infty}
\mathbb{E}\!\left[
\langle \sigma_A\sigma_B\rangle_n^2
\right]
=0 .
\ee
Hence there is no spin-glass order at any finite temperature.
\end{theorem}
%%%%%%%%%%%%%%%%%%%%%%%%%%%%%%%%%%%%%%%%%%%%%%%%%%%%%%%%%%%%%%%
%%%%%%%%%%%%%%%%%%%%%%%%%%%%%%%%%%%%%%%%%%%%%%%%%%%%%%%%%%%%%%%

Condition~\eqref{sufficient-condition} has several immediate consequences. 
For fixed \(s\), the right-hand side of condition~\eqref{sufficient-condition} increases with \(b\). Since the fractal dimension $d$ decreases as \(b\) increases, the no-order criterion becomes easier to satisfy at lower fractal dimensions.

For \(s=b\), corresponding to \(d=2\), condition~\eqref{sufficient-condition} is satisfied for any even \(s\). 
This case includes the MK approximation to the Ising spin glass on the square lattice, and Theorem~1 therefore gives a rigorous proof of the absence of spin-glass order within this approximation.
By contrast, for \(s=b^2\), corresponding to \(d=3\), condition~\eqref{sufficient-condition} is not satisfied for any even \(b\). This is consistent with previous studies indicating the existence of a spin-glass phase in three dimensions within the MK approximation~\cite{SY}. We emphasize, however, that condition~\eqref{sufficient-condition} is only sufficient: when it fails, our method does not imply the presence of spin-glass order.

For example, condition~\eqref{sufficient-condition} is satisfied for \((s,b)=(6,5)\) and \((s,b)=(8,6)\), corresponding to \(d\simeq2.11\) and \(d\simeq2.16\), respectively. 
On the other hand, it is not satisfied for \((s,b)=(6,3)\), corresponding to \(d\simeq2.63\).
This pattern is consistent with the commonly quoted estimate \(d_{\rm lc}^{\rm MK}\simeq2.52\), which was inferred from numerical studies of MK hierarchical lattices with relatively small values of \(s\) and \(b\)~\cite{DTB}.

However, the criterion is not sharp. 
For instance, condition~\eqref{sufficient-condition} is not satisfied for \((s,b)=(4,3)\), \((s,b)=(6,4)\), or \((s,b)=(8,5)\), although these lattices have fractal dimensions \(d\simeq2.26\), \(d\simeq2.29\), and \(d\simeq2.29\), respectively.
Previous studies suggest the absence of a spin-glass phase in such cases~\cite{DTB}, but the present one-step dilution criterion is not strong enough to establish this absence.
One possible way to improve the criterion is to perform two or more exact RG steps before making the percolation comparison, thereby increasing the probability that the effective interaction vanishes. 
This refinement is useful in the analysis of the self-dual hierarchical lattice discussed below.

The same condition leads to a more striking consequence.
For large even \(s\), Stirling's formula gives $p_{s,b} \simeq \sqrt{{2}/{(\pi s)}}$.
We then choose \(b\) to be of the form $b\simeq
\sqrt{{s\pi}/{2}}\bigl(\log s-\log\log s+C\bigr)$,
where \(C\) is a constant independent of \(s\). Any fixed
\(C>\log 2\) ensures that condition~\eqref{sufficient-condition}
is satisfied for all sufficiently large even \(s\).
The corresponding fractal dimension is
\be
d
=
1+\frac{\log s}{\log b}
\simeq
3-
\frac{4\log\log s+O(1)}{\log s}.
\ee
Thus, by taking \(s\) and \(b\) sufficiently large, one can construct MK hierarchical lattices that satisfy the no-order criterion even though their fractal dimensions are arbitrarily close to three from below. 
This contrasts with the numerical estimate \(d_{\rm lc}^{\rm MK}\simeq2.52\)~\cite{DTB}, obtained from MK hierarchical lattices with relatively small scale factors and branching numbers. 
Our result therefore shows that, within the family of MK hierarchical lattices, the lower-critical behavior of Ising spin glasses depends not only on the fractal dimension but also on the detailed hierarchical construction.

A simple scaling argument~\cite{Boettcher} offers a heuristic interpretation. 
For sufficiently large \(b\), a single RG step reduces the characteristic interaction scale along each branch by a factor of \(1/b\), while summing the approximately independent contributions from \(s=b^{d-1}\) branches enhances it by a factor of \(\sqrt{s}\). 
The net rescaling factor is therefore ${\sqrt{s}}/{b}=b^{(d-3)/2}$, so that \(d=3\) is marginal in the large-\(s\), large-\(b\) regime. 
Our proof rigorously establishes the no-order side of this scaling argument.
This scaling reflects a connectivity effect specific to the MK hierarchical construction and should not be interpreted as a direct statement about regular lattices in Euclidean space, for which the lower critical dimension is estimated to be \(d_{\rm lc}\simeq2.5\)~\cite{MP,Boettcher,FPV}.

We next consider the zero-temperature limit. At \(T=0\), the exact RG equation~\cite{Nobre} is
\be
J_{LR}
=
\sum_{\alpha=1}^{s}
\operatorname{sgn}
\left(
\prod_{k=0}^{b-1}J_{\alpha k}
\right)
\min_{0\le k\le b-1}|J_{\alpha k}| .
\ee
For the symmetric binary distribution, each branch contribution again has the same magnitude and an independent random sign.
Therefore, for even \(s\), the probability that the effective interaction vanishes after one RG step is again \(p_{s,b}\). The same percolation argument then gives the zero-temperature counterpart of Theorem~1.

%%%%%%%%%%%%%%%%%%%%%%%%%%%%%%%%%%%%%%%%%%%%%%%%%%%%%%%%%%%%%%%
\begin{theorem}
Let \(s\) be even and suppose that condition~\eqref{sufficient-condition} holds.
Then there exist constants \(C>0\) and \(c>0\), independent of \(n\),
such that $\lim_{\beta\to\infty}
\mathbb{E}\left[
\langle \sigma_A\sigma_B\rangle_n^2
\right]
\leq C\exp(-cr_n)$.
In particular,
\begin{equation}
\lim_{n\to\infty}\lim_{\beta\to\infty}
\mathbb{E}\left[
\langle \sigma_A\sigma_B\rangle_n^2
\right]
=0.
\end{equation}
Thus there is no spin-glass order even at zero temperature.
\end{theorem}
%%%%%%%%%%%%%%%%%%%%%%%%%%%%%%%%%%%%%%%%%%%%%%%%%%%%%%%%%%%%%%%
%%%%%%%%%%%%%%%%%%%%%%%%%%%%%%%%%%%%%%%%%%%%%%%%%%%%%%%%%%%%%%%
The same exponential bound implies that, in this regime, the zero-temperature correlation length is finite.
The system is therefore noncritical and lies in the paramagnetic regime even at zero temperature, in agreement with previous numerical studies~\cite{Gingras,ANC}. 
This conclusion relies crucially on the discreteness of the binary coupling distribution and on the branching number being even.
For odd \(s\) or continuous coupling distributions such as the Gaussian distribution, systems below the lower critical dimension are instead generally expected to be critical at zero temperature: the correlation length diverges, and the low-temperature behavior is governed by a zero-temperature fixed point~\cite{JK}.

We finally discuss stiffness. 
At zero temperature, stiffness is characterized by the boundary-condition energy difference \(\Delta E_n=E_n^{++}-E_n^{+-}\); at finite temperature, the analogous
quantity is \(\Delta F_n=F_n^{++}-F_n^{+-}\). 
Here, the superscripts $(++)$ and $(+-)$ denote the boundary conditions $(\sigma_A,\sigma_B)=(1,1)$ and $(1,-1)$, respectively.
If \(|\Delta E_n|\sim r_n^\theta\), \(\theta\) is the stiffness exponent.
A positive value, \(\theta>0\), indicates that large-scale domain-wall excitations become increasingly costly and is commonly regarded as evidence for a stable spin-glass phase.
Below the lower critical dimension, Gaussian couplings typically yield \(\theta<0\), whereas binary couplings exhibit an apparent pinning near
zero because of ground-state degeneracy~\cite{AMMP}.

In the present setting, our dilution argument gives a stronger conclusion. 
If no path of nonzero effective interactions connects \(A\) and \(B\), changing the boundary condition at \(B\) cannot affect the component containing \(A\). 
By spin-flip symmetry, the component containing \(B\) has the same energy at zero temperature, or the same free energy at finite temperature, for \(\sigma_B=1\) and \(\sigma_B=-1\).
Thus, on the disconnection event, \(\Delta E_n=0\) at zero temperature and \(\Delta F_n=0\) at finite temperature.
Combining this observation with Theorems~1 and 2 gives the following result.

%%%%%%%%%%%%%%%%%%%%%%%%%%%%%%%%%%%%%%%%%%%%%%%%%%%%%%%%%%%%%%%%
%%%%%%%%%%%%%%%%%%%%%%%%%%%%%%%%%%%%%%%%%%%%%%%%%%%%%%%%%%%%%%%
\begin{theorem}
Let \(s\) be even and suppose that condition~\eqref{sufficient-condition} holds. 
Then there exist constants \(C>0\) and \(c>0\), independent of \(n\), such that, at zero temperature, $\Pr[\Delta E_n=0]
\ge
1-Ce^{-c r_n}$,
and at finite temperature, $\Pr[\Delta F_n=0]
\ge
1-Ce^{-c r_n}$.
Therefore, the boundary-condition response vanishes in probability in the thermodynamic limit, and the system has no stiffness at any temperature.
\end{theorem}
%%%%%%%%%%%%%%%%%%%%%%%%%%%%%%%%%%%%%%%%%%%%%%%%%%%%%%%%%%%%%%%
This result is stronger than the apparent \(\theta\simeq0\) behavior often observed for discrete coupling distributions below the lower critical dimension~\cite{AMMP}.  
In the present regime, the probability that the boundary-condition energy difference vanishes at zero temperature, or that the corresponding free-energy difference vanishes at finite temperature, approaches one exponentially fast as the endpoint distance increases.
Thus, the conventional power-law definition of a stiffness exponent is not appropriate. Formally, one may regard the behavior as corresponding to \(\theta=-\infty\), rather than to genuine power-law scaling with \(\theta=0\). 
This extreme behavior appears to be a special consequence of combining a discrete coupling distribution with an even branching number and is not expected to persist when the branching number is odd or the coupling distribution is continuous.

%%%%%%%%%%%%%%%%%%%%%%%%%%%%%%%%%%%%%%%%%%%%%%%%%%%%%%%%%%%%%%%
%\paragraph*{Other probability distributions and odd \(s\)--}
\paragraph*{Extensions and limitations of the dilution criterion---}
So far, we have focused on the symmetric binary distribution. The same dilution argument can be extended to an asymmetric binary distribution, $P(J_{ij})
=
\rho \delta(J_{ij}-1)
+
(1-\rho)\delta(J_{ij}+1)
\quad$ $(0\le \rho \le 1)$.
For even \(s\), the probability that the renormalized interaction vanishes after one RG step is
\be
p_{s,b}(\rho)
=
\binom{s}{s/2}
\left[
\frac{1-(2\rho-1)^{2b}}{4} 
\right]^{s/2}.\label{psb-as}
\ee
Unlike in the symmetric case, this probability depends on both \(s\) and \(b\). 
%For \(\rho=1/2\), Eq.~\eqref{psb-as} reduces to Eq.~\eqref{psb}, while for \(\rho=0\) or \(\rho=1\) it gives \(p_{s,b}(\rho)=0\), as expected.
The percolation criterion itself is unchanged. Namely, if $p_{s,b}(\rho)
<
\left[1-\{1-p_{s,b}(\rho)\}^b\right]^s$,
then the induced dilution exceeds the corresponding percolation threshold, and the endpoint connectivity probability decays exponentially with the distance between the endpoints.
Consequently, both spin-glass order and stiffness are absent at all temperatures. 
Moreover, in the asymmetric case, the same argument also rules out ferromagnetic long-range order as diagnosed by the endpoint correlation, because
\(\left|\mathbb{E}[\langle \sigma_A\sigma_B\rangle_n]\right|\)
is bounded by the probability that the two endpoints are connected through nonzero effective interactions.

The argument also applies, in principle, to more general discrete coupling distributions whenever a single RG step produces an atom at \(J_{LR}=0\) with probability \(p>0\). In such cases, the same percolation comparison gives a sufficient condition for the absence of spin-glass order, with \(p\) replacing \(p_{s,b}\). 

By contrast, the present argument does not apply to continuous coupling distributions such as the Gaussian distribution. 
The percolation comparison requires an atom of positive mass at \(J_{LR}=0\), because only bonds that vanish exactly can be removed without changing the Hamiltonian. 
For a continuous distribution, exact cancellation occurs with probability zero, so the induced vacancy probability is zero. 
A similar obstruction arises for binary couplings when \(s\) is odd: an odd number of equal-magnitude branch contributions cannot cancel exactly. 
The one-step criterion therefore yields no conclusion in either case.

%%%%%%%%%%%%%%%%%%%%%%%%%%%%%%%%%%%%%%%%%%%%%%%%%%%%%%%%%%%%%%%
\paragraph*{Absence of spin-glass order on a self-dual hierarchical lattice---}
We next apply the same dilution mechanism to the Ising spin-glass model on the self-dual hierarchical lattice, whose fractal dimension is $d=\log 5/\log 2\simeq 2.32$.
The generating unit consists of two external spins, \(\sigma_L\) and \(\sigma_R\), two internal spins, \(\sigma_1\) and \(\sigma_2\), and five bonds. 
Its Hamiltonian is
\be
\mathcal H_{\rm unit}&=&-J_{L1}\sigma_L\sigma_1- J_{1R}\sigma_1\sigma_R -  J_{L2}\sigma_L\sigma_2- J_{2R}\sigma_2\sigma_R
\no\\
&& - J_{12}\sigma_1\sigma_2.
\ee
Compared with the generating unit of the MK hierarchical lattice with \(s=b=2\), the present generating unit contains an additional bond \(J_{12}\) connecting the two internal spins.
By summing exactly over the internal spins \(\sigma_1\) and \(\sigma_2\), we obtain an effective interaction between \(\sigma_L\) and \(\sigma_R\).
The explicit RG equations at finite and zero temperature are given in the End Matter.
Here we use only the vacancy probabilities resulting from these exact RG steps.

On the self-dual hierarchical lattice, let \(q\) denote the probability that a given effective bond is absent. For a single generating unit, the probability that no path connects the two endpoints is $\Psi(q)=q^2(2+2q-5q^2+2q^3)$.
Thus, if $q<\Psi(q)$, the vacancy probability increases under successive iterations of the hierarchical percolation recursion. 
In this case, the absence of spin-glass order and stiffness follows from the same connectivity argument used in Theorem~1.
The map \(q\mapsto\Psi(q)\) has an unstable fixed point at \(q=1/2\), and any initial value \(q>1/2\) flows to \(q=1\) under iteration.
It therefore remains to show that the vacancy probability exceeds \(1/2\) after finitely many exact RG steps.

For the symmetric binary distribution, direct enumeration using the exact RG equations gives \(q_1=1/2\) after one RG step, at every finite temperature and at zero temperature. This value is exactly the percolation threshold and is therefore not sufficient by itself. 
We therefore perform a second exact RG step. 
The proof of Theorem~4 in the End Matter gives \(q_2=261/512\) at finite temperature and \(q_2=2255/4096\) at \(T=0\).
Both are strictly larger than \(1/2\). 
Hence, the vacancy probability flows to one under further hierarchical iterations, and the same connectivity argument used in Theorem~1 proves the absence of spin-glass order and stiffness.

%%%%%%%%%%%%%%%%%%%%%%%%%%%%%%%%%%%%%%%%%%%%%%%%%%%%%%%%%%%%%%%
\begin{theorem}
In the thermodynamic limit, the Ising spin glass with symmetric binary couplings on the self-dual hierarchical lattice with \(d=\log 5/\log 2\) exhibits neither spin-glass order nor stiffness at any temperature, including zero temperature.
\end{theorem}
%%%%%%%%%%%%%%%%%%%%%%%%%%%%%%%%%%%%%%%%%%%%%%%%%%%%%%%%%%%%%%%

%%%%%%%%%%%%%%%%%%%%%%%%%%%%%%%%%%%%%%%%%%%%%%%%%%%%%%%%%%%%%%%
\paragraph*{Conclusions---}

Rigorous analysis of finite-dimensional spin glasses remains a challenging problem. 
In this Letter, we addressed this problem through an exact real-space RG analysis on hierarchical lattices. We introduced a rigorous dilution criterion for excluding spin-glass order. 
For symmetric binary couplings and an even branching number, a single exact RG step creates zero effective bonds with positive probability.
When this induced dilution exceeds the corresponding hierarchical percolation threshold, the endpoint connectivity probability decays exponentially with distance.
Consequently, both spin-glass order and stiffness are absent at all temperatures, including zero temperature.

This criterion proves the absence of spin-glass order and stiffness in the Ising spin-glass model on the square lattice within the MK approximation. 
In the asymptotic large-\(s\), large-\(b\) regime, the criterion applies to families of MK hierarchical lattices whose fractal dimensions approach three from below, thereby providing rigorous support for earlier scaling arguments identifying \(d=3\) as the marginal dimension~\cite{Boettcher}.
Because this result emerges only for sufficiently large \(s\) and \(b\), it does not conflict with the commonly quoted estimate \(d_{\rm lc}^{\rm MK}\simeq2.52\), obtained from MK hierarchical lattices with relatively small \(b\) and \(s\). 
Rather, it shows that the lower-critical behavior within the family of MK hierarchical lattices depends sensitively on the detailed lattice construction and not on the fractal dimension alone. 
The numerical proximity of the MK estimate to \(d_{\rm lc}\simeq2.5\) for regular lattices in Euclidean space~\cite{MP,Boettcher,FPV} should not be interpreted as evidence for a simple quantitative correspondence between the two settings.

This mechanism also proves the absence of endpoint spin-glass order and stiffness on the self-dual hierarchical lattice with \(d=\log 5/\log 2\).
Several important cases remain open, including the case of an odd branching number, continuous coupling distributions, and even-branching MK lattices for which the sufficient condition is not satisfied.

An important further problem is whether the present method can be extended to spin-glass models in two dimensions, where no spin-glass phase is expected at finite temperature. 
Previous work has provided analytical evidence for the absence of such a phase on both the self-dual hierarchical lattice studied here and the square lattice by combining graph duality, the replica method, and real-space RG arguments~\cite{ON}.
The approach based on exact dilution may provide a useful new route toward a rigorous understanding of this problem.

%%%%%%%%%%%%%%%%%%%%%%%%%%%%%%%%%%%%%%%%%%%%%%%%%%%%%%%%%%%%%%%%%%
This work was supported by JST BOOST, Japan (Grant No. JPMJBY24B6), and JSPS KAKENHI (Grant No. 24K16973).
This work was was supported by the Cross-ministerial Strategic Innovation Promotion Program (SIP) of the Cabinet Office (No. 23836436).

%%%%%%%%%%%%%%%%%%%%%%%%%%%%%%%%%%%%%%%%%%%%%%%%%%%%%%%%%%%%%
%%%%%%%%%%%%%%%%%%%%%%%%%%%%%%%%%%%%%%%%%%%%%%%%%%%%%%%%%%%%%%%%%%%%%%%%%%%%

%%%%%%%%%%%%%%%%%%%%%%%%%%%%%%%%%%%%%%%%%%%%%%%%%%%%%%%%%%%%%%%%%%%%%%%%%%%
%%%%%%%%%%%%%%%%%%%%%%%%%%%%%%%%%%%%%%%%%%%%%%%%%%%%%%%%%%%%%%%%%%%%%%%%%%%
\section{End Matter}

%%%%%%%%%%%%%%%%%%%%%%%%%%%%%%%%%%%%%%%%%%%%%%%%%%%%%%%%%%%%%%%
\paragraph*{Proof of Theorem 1---}
We prove Theorem~1.
After one exact RG step, each effective bond in the reduced hierarchical lattice is absent with probability \(p_{s,b}\).
We use \(p_{s,b}\) as the initial vacancy probability for the percolation recursion.

Let \(y_n\) denote the probability that no path of nonzero effective interactions connects the endpoints \(A\) and \(B\) in a generation-\(n\) MK hierarchical lattice.
Then, by the hierarchical construction,
\be
y_n
\ge
\left[1-(1-y_{n-1})^b\right]^s,
\qquad
y_1=p_{s,b}.
\ee
Equivalently, in terms of the connectivity probability \(x_n=1-y_n\), we have
\be
x_n
\le
f(x_{n-1}),
\qquad
x_1=1-p_{s,b},
\ee
where
\be
f(x)=1-\left(1-x^b\right)^s .
\ee

The function \(f\) is strictly increasing on \([0,1]\). 
We next show that \(f\) has a unique nontrivial fixed point in \((0,1)\) for \(s>1\).
Indeed, a nontrivial fixed point \(x\in(0,1)\) satisfies
\be
x=1-(1-x^b)^s ,
\ee
or equivalently
\be
R_b(x):=
\frac{\log(1-x)}{\log(1-x^b)}
=s .
\ee
The function \(R_b\) is strictly decreasing on \((0,1)\).
Moreover,
\be
\lim_{x\downarrow0}R_b(x)=\infty,
\qquad
\lim_{x\uparrow1}R_b(x)=1 .
\ee
Hence, for \(s>1\), the equation \(R_b(x)=s\) has exactly one solution in \((0,1)\). 
Thus, \(f\) has exactly one nontrivial fixed point in \((0,1)\), in addition to the trivial fixed points \(0\) and \(1\).

Condition~\eqref{sufficient-condition} is equivalent to
\be
f(x_1)<x_1 .
\ee
Therefore, \(x_1\) lies below the unique nontrivial fixed point. 
Since \(f(x)<x\) below the nontrivial fixed point and \(f\) is increasing, the sequence \((x_n)\) decreases monotonically to zero.
Since the endpoint spin correlation vanishes on the disconnection event and is bounded in absolute value by one otherwise, we have
\be
\mathbb{E}\!\left[
\langle \sigma_A\sigma_B\rangle_n^2
\right]
\le x_n .
\ee
It follows that
\be
\lim_{n\to\infty}
\mathbb{E}\!\left[
\langle \sigma_A\sigma_B\rangle_n^2
\right]
\le
\lim_{n\to\infty}x_n
=0 .
\ee

%%%%%%%%%%%%%%%%%%%%
%%%%%%%%%%%%%%%%%%%%
%%%%%%%%%%%%%%%%%%%%

It remains to estimate the rate of decay. 
Since \(x_n\to0\), we can choose \(n_0\) sufficiently large so that
\be
\lambda:=s^{1/(b-1)}x_{n_0}<1 .
\ee
Using
\be
f(x)=1-\left(1-x^b\right)^s\le s x^b ,
\ee
we obtain, for all \(n\ge n_0\),
\be
x_n
\le
s^{\frac{b^{n-n_0}-1}{b-1}}
x_{n_0}^{b^{n-n_0}}
=
s^{-\frac{1}{b-1}}
\left(
s^{1/(b-1)}x_{n_0}
\right)^{b^{n-n_0}} .
\ee
Thus,
\be
x_n
\le
C_0\lambda^{b^{n-n_0}},
\qquad
C_0=s^{-1/(b-1)} .
\ee
Since \(r_n=b^n\), this bound can be written as
\be
x_n
\le
C e^{-c r_n}
\ee
for some constants \(C>0\) and \(c>0\). Therefore,
\be
\mathbb{E}\!\left[
\langle \sigma_A\sigma_B\rangle_n^2
\right]
\le x_n
\le
C e^{-c r_n},
\ee
which proves Theorem~1.

%%%%%%%%%%%%%%%%%%%%%%%%%%%%%%%%%%%%%%%%%%%%%%%%%%%%%%%%%%%%%%%
\paragraph*{Proof of Theorem 4---}
We first record the exact RG equations used in the direct enumeration.
For the self-dual generating unit defined in the main text, summing over
the two internal spins gives
\begin{equation}
t_{LR}
=
\frac{
t_{L1}t_{1R}
+t_{L1}t_{12}t_{2R}
+t_{L2}t_{12}t_{1R}
+t_{L2}t_{2R}
}{
1
+t_{L1}t_{L2}t_{12}
+t_{12}t_{1R}t_{2R}
+t_{L1}t_{L2}t_{1R}t_{2R}
},
\end{equation}
where \(t_{ij}=\tanh(\beta J_{ij})\) and
\(t_{LR}=\tanh(\beta J_{LR})\). At zero temperature, the corresponding
exact RG equation~\cite{Nobre} is
\begin{align}
J_{LR}
={}&
\frac{1}{2}
\max\Bigl(
J_{12}
+\left|J_{L1}+J_{L2}+J_{1R}+J_{2R}\right|,
\nonumber\\
&\hspace{2.8cm}
-J_{12}
+\left|J_{L1}-J_{L2}+J_{1R}-J_{2R}\right|
\Bigr)
\nonumber\\
&-
\frac{1}{2}
\max\Bigl(
J_{12}
+\left|J_{L1}+J_{L2}-J_{1R}-J_{2R}\right|,
\nonumber\\
&\hspace{2.8cm}
-J_{12}
+\left|J_{L1}-J_{L2}-J_{1R}+J_{2R}\right|
\Bigr).
\end{align}

For bond percolation on the self-dual hierarchical lattice, the vacancy
probability transforms as
\begin{equation}
\Psi(q)=q^2(2+2q-5q^2+2q^3).
\end{equation}
Since
\begin{equation}
\Psi(q)-q=q(q-1)(2q-1)(q^2-q-1),
\end{equation}
we have \(\Psi(q)>q\) for \(1/2<q<1\). 
Hence, once \(q>1/2\), the vacancy probability flows to one under further hierarchical iterations.

We now summarize the direct enumeration. 
For the symmetric binary distribution, the RG equation at finite temperature yields
\be
t_{LR}=0
\ee
for \(16\) of the \(2^5\) bond configurations.
The remaining configurations yield four nonzero values,
\be
t_{LR}=\pm t_+,\ \pm t_-,
\ee
each occurring with probability \(1/8\), where
\begin{equation}
t_+
=
\frac{2t^2(1+t)}{1+2t^3+t^4},
\qquad
t_-
=
\frac{2t^2(1-t)}{1-2t^3+t^4},
\qquad
t=\tanh\beta .
\end{equation}
Thus, the vacancy probability after one RG step is
\begin{equation}
q_1=\frac{1}{2}.
\end{equation}
A direct enumeration of the second RG step, using the above five-point distribution, gives
\begin{equation}
q_2=\frac{261}{512}>\frac{1}{2}.
\end{equation}
Therefore, at any finite temperature, two RG steps produce a vacancy probability that exceeds the percolation threshold.

At zero temperature, the first RG step yields
\be
J_{LR}=0,\ \pm1,\ \pm2
\ee
with probabilities
\be
\frac{1}{2},\quad
\frac{1}{8},\quad
\frac{1}{8},\quad
\frac{1}{8},\quad
\frac{1}{8},
\ee
respectively. 
Applying the zero-temperature RG equation once more yields
\begin{equation}
q_2=\frac{2255}{4096}>\frac{1}{2}.
\end{equation}
Thus, the vacancy probability again exceeds the threshold after two RG steps.
This proves the absence of endpoint spin-glass order and stiffness on the self-dual hierarchical lattice.

\end{document}